\documentclass[12pt,draftcls,onecolumn,journal]{IEEEtran}
\usepackage{amsmath,amsthm}
\usepackage{amsfonts}%
\usepackage{amssymb}%
\usepackage{graphicx}
\usepackage{epsfig}
\usepackage{setspace}
\usepackage{cite}
\usepackage[pagewise]{lineno}

\newtheorem{thm}{Theorem}[section]
\newtheorem{exmp}{Example}[section]
\newtheorem{lem}{Lemma}[section]
\newtheorem{prop}{Proposition}[section]
\numberwithin{equation}{section}

\newcommand{\sysref}[1]{system~(\ref{#1})}

\newcommand*{\figref}[1]{{ Fig.~\ref{#1}}}
\newcommand{\thmref}[1]{{\bf Theorem~\ref{#1}}}
\newcommand{\lmref}[1]{{\bf Lemma~\ref{#1}}}
\newcommand{\propref}[1]{ {\bf Proposition~\ref{#1}}}
\newcommand*{\tabref}[1]{{\sc Table~\ref{#1}}}

\usepackage{mdwmath}
\usepackage{mdwtab}

\begin{document}
\title{Kovalenko's Full-Rank Limit and Overhead as Lower Bounds for Error-Performances of  LDPC and LT Codes over Binary Erasure Channels}

\author{
Ki-Moon Lee,
        Hayder Radha,~\IEEEmembership{Senior~Member,~IEEE,} and 
        Beom-Jin Kim% <-this % stops a space
\thanks{Ki-Moon Lee is with 
Communication and Coding Theory Lab, Dept of Information and Communication Engineering, Sungkyunkwan University, Suwon Kyungki, Korea.   e-mail:leekimoo@skku.edu}% <-this % stops a space
\thanks{Hayder Radha is with Dept of Electrical and Computer Engineering, Michigan State University,  E. Lansing, MI-48824, USA.  e-mail:radha@egr.msu.edu}
\thanks{Beom-Jin Kim is with Dept of Mathematics Natural Science, Yonsei University, Seoul Korea.  e-mail:beomjinkim@yonsei.ac.kr}

% <-this % stops a space
%\thanks{A short version of this paper was accepted to ISITA 2008, Auckland NZ.  The first draft was submitted to IEEE Transactions on Information Theory, 2008/06.}
}


%\markboth{Journal of \LaTeX\ Class Files,~Vol.~6, No.~1, January~2007}%
%{Shell \MakeLowercase{\textit{et al.}}: Bare Demo of IEEEtran.cls for Journals}
\markboth{2nd Draft}%
{Ki-Moon Lee \MakeLowercase{\textit{et al.}}}



\maketitle

%\begin{abstract}
%%\boldmath
%The abstract goes here.
%\end{abstract}


\begin{abstract}
We present  Kovalenko's full-rank limit as  a tight lower bound  for decoding error probability of  LDPC  codes and LT  codes over BEC. 
From the limit, we derive a full-rank overhead as a lower bound  for stable overheads for successful maximum-likelihood decoding of the codes. 
%We also provide that, when a number of dense rows are supplemented to check matrices of codes, the DEP of the  codes can be predicted tightly by using a union bound.  
\end{abstract}

%\begin{IEEEkeywords}
%Kovalenko's Full-Rank Limit and Overhead, Error-Performance, LT Codes, LDPC Codes, Binary Erasure Channels.
%\end{IEEEkeywords}

%
%\begin{IEEEkeywords}
%IEEEtran, journal, \LaTeX, paper, template.
%\end{IEEEkeywords}

\section{Introduction and Backgrounds \label{intro}}
Binary Erasure Channels (BEC) based Low-Density Parity-Check (LDPC) codes \cite{eeldpc,bnm1} and Luby Transform (LT) codes \cite{lt,raptor} became quite popular for a variety of applications over packet networks such as the Internet.
The popularity of LDPC and LT codes are due in part to (a) the low-complexity of the popular set of decoding algorithms that fall under the umbrella of the Message Passing Algorithm (MPA) (otherwise called Belief Propagation Algorithm for BEC) \cite{eeldpc,bnm1}, (b) good error performance of MPA for codes of large block lengths, and (c) the flexibility in choosing the block lengths of these codes, which make them usable for a variety of applications.

In BEC, without loss of generality, the task of both LT and LDPC decoders is to recover the unique solution of  a consistent linear system
\begin{equation}
HX^T=\beta^T, \quad \beta=(\beta_1,\dots,\beta_m)\in (\mathbb{F}_2^s)^m, \label{sys:ini}
\end{equation} 
where $H$ is an $m\times n$ matrix over $\mathbb{F}_2$.  
This can be explained shortly as follows.
In case of LT codes, to communicate an information symbol vector  $\alpha=(\alpha_1,\dots,\alpha_n) \in (\mathbb{F}_2^s)^n$, a sender constantly generates and transmits  a syndrome symbol  $\beta_i=H_i\alpha^T$ over BEC, where   $H_i\in \mathbb{F}_2^n$ is generated uniformly at random  on the fly by using the Robust Soliton Distribution $\mu(x)=\sum \mu_d x^d$ (see \cite{lt}).   A receiver then acquires  a set of pairs $\{ (H_{i_t},\beta_{i_t})\}_{t=1}^{m}$ and interprets it as System~(\ref{sys:ini}). Hence,  the variable vector  $X=(x_1,\dots,x_n)\in (\mathbb{F}_2^s)^n$ in the system  represents the information symbol vector $\alpha$.
 In case of LDPC codes, contrastingly,  a sender transmits a codeword vector $\alpha=(\alpha_1,\dots,\alpha_N)$ in $\text{Ker}(M)=\{\alpha \in (\mathbb{F}_2^s)^N \:|\: M \cdot \alpha^T=0  \}$, where $M$ is an $m\times N$ binary check matrix. Due to erasures, some of symbols of $\alpha$ may be lost and a receiver acquires a part of $\alpha$, denote it as $\bar\alpha$.   Then by the rearrangements  $\alpha\equiv (\bar{\alpha},X)$ and  $M\equiv [\bar{H};H]$,  where $\bar{H}$ and $H$ consist of columns of $M$ that associate symbols of $\bar\alpha$ and $X$, respectively,  the receiver interprets the kernel space constraint $M\cdot \alpha^T=0$ as System~(\ref{sys:ini}),  where $\beta^T=\bar{H}\bar{\alpha}^T$. Hence in LDPC codes, $X$  represents a lost symbol vector of $\alpha$.

In LT codes,  the column-dimension $n$ of $H$ is fixed, the row-dimension $m$ of $H$ is a variable, and a reception overhead  $\gamma=\frac{m-n}{n}$ is the key parameter for measuring error-performance of codes.
  In LPDC codes, however, the row-dimension $m$  is fixed in general,  the column-dimension $n=pN$  is a variable, and  a erasure rate (or loss rate) $p=\frac{n}{N}$ is the key parameter for measuring error-performance of codes.   
    Let $R=1-\frac{m}{N}$, a code-rate of an LDPC code.  By using $m=(1+\gamma)n$, $n=pN$, and $R=1-\frac{m}{N}$, $p$ and $\gamma$ are expressed as
 \begin{equation}
 p=\frac{1-R}{1+\gamma}\quad \text{and}\quad  \gamma=\frac{m-n}{n}=\frac{1-(R+p)}{p}. \label{eq:gammap}
 \end{equation}
Like LT codes, thus, the error-performance of LDPC codes  can be also measured in terms of $\gamma$.

 Several literatures showed the existence of capacity approaching LDPC codes \cite{ca1} and optimal LT codes \cite{lt,raptor}, whose minimal overheads  for successful decoding by the MPA in high probability tends to zero as  block lengths ($n$ for LT and $N$ for LDPC codes) increase to infinity.    For  codes of short block lengths,  however,  their minimum overheads  (for the successful decoding by the MPA in high probability) are not close to zero.  Furthermore, even for a nontrivial $\gamma>0$, the full-rank probability $\Pr(\text{Rank}(H)=n)$ is not very close to $1$.

System~(\ref{sys:ini}) has its unique solution, iff, $\text{Rank}(H)=n$ the full rank of $H$. In case of  the full-rank, the unique solution can be recovered  by using a Maximum-Likelihood Decoding Algorithm (MLDA) such as the ones in \cite{bnm1,ourlt,shortlt,usp1}.  These algorithms are an efficient  Gaussian Elimination (GE) that fully utilize an approximate lower triangulation of $H$, which is obtainable by using the diagonal extension process with various greedy algorithms \cite{eeldpc,bnm1}.    
%The extension process also embeds the MPA automatically.  
  Under those GE based MLDAs, thus, the probability of decoding success  is precisely the  $\Pr(\text{Rank}(H)=n)$.
Let us  define the Decoding Error Probability (DEP) of a code as the rank-deficient probability 
\begin{equation}
P^{err}_{ML}(1+\gamma,n)=1-\Pr(\text{Rank}(H)=n),  \label{eq:dfr}
\end{equation}
where $H$ is an $m\times n$ decoder matrix of \sysref{sys:ini} with $\gamma=\frac{m-n}{n}$.
Assume that $P^{err}_{ML}(1+\gamma,n)$ is a decreasing function with respect to $\gamma$.
Then for a given  error-bound (or deficiency bound) $0\le \delta\le 1$,  define 
\begin{equation}
\gamma_*(\delta,n) = \min_{\gamma \ge 0}\{ \gamma\: |\: P^{err}_{ML}(1+\gamma,n)<\delta \}, \label{eq:mo}
\end{equation}
and refer to as the Minimum Stable Overhead (MSO) of a code within the error-bound  $\delta$.
Since  $P^{err}_{ML}(1+\gamma,n)$ is decreasing, we may expect that $P^{err}_{ML}(1+\gamma,n) \le \delta$ for any $\gamma \ge \gamma_*(\delta,n)$.    
%We refer to such $\gamma$ as a stable overhead within error-bound $\delta$.
Thus, the key part of designing codes is  to identify lower bounds of DEP and MSO then to obtain  the   codes whose DEP and MSO are close to the bounds.

 In this paper, as the main contribution of this paper, we define Kovalenko's Full-Rank Limit (KFRL), denote as $K(1+\gamma,n)$, from Kovalenko's rank-distribution of binary random matrices  \cite{kv1,cc,rg}, and show that  it is a probabilistic lower bound for $P^{err}_{ML}(1+\gamma,n)$, i.e., $K(1+\gamma,n)\le P^{err}_{ML}(1+\gamma,n)$ for any $\gamma$ and $n$.  
 We then derive  Kovalenko's Full-Rank Overhead (KFRO) from KFRL, denote as $\gamma_K(\delta,n)$, as a lower-bound for  MSO, i.e., $\gamma_K(\delta,n)\le \gamma_*(\delta,n)$ for any $\delta$ and $n$, and show that the overhead $\gamma_K(\delta,n)$ tells the least number of symbols that  a receiver should acquire to achieve $P^{err}_{ML}(1+\gamma,n)\le \delta$.  
 We also provide experimental evidences  which  show the viability that, given a destined error-bound $\delta_0$,  both LT and LDPC codes may be designed to achieve their error-performances in $P^{err}_{ML}(1+\gamma,n)$ and $\gamma^*(\delta,n)$ that are close to  $K(1+\gamma,n)$ and  $\gamma_K(\delta,n)$ for  $\delta \ge \delta_0$, respectively,  by supplementing enough number of dense rows to $H$ of \sysref{sys:ini}.
 
% We also provide experimental results which show that, both $P^{err}_{ML}(1+\gamma,n)$ and $\gamma_*(\delta,n)$ may approach to $K(1+\gamma,n)$ and $\gamma_K(\delta,n)$, respectively, when enough number of dense rows (or columns) are supplemented to $H$.
 
 The remainder of this paper is composed of as follows.
%What we shall focus on in the remainder of the paper are as follows.  
In Section~\ref{kfro}, we define KFRL and KFRO and verify them as lower bounds for DEP and MSO of LDPC and LT codes. In Section~\ref{simulations}, we present experimental results of the performances of codes in terms of DEP and overhead.   We summarize the paper in Section~\ref{summary}.

\section{Kovalenko's Full-Rank Limit and Overheads \label{kfro}}

Let us first clarify terms and notations for the remainder of this section. Let $|H_i|$ denote the number of nonzero entries of a row $H_i$ of $H$ and refer to as the degree of $H_i$.  
Given an overhead $\gamma$, we shall assume that $\gamma n=k$ for some integer $k\ge 0$.
Let $\hat{H}$ denote an  $m\times n $ random binary matrix over $\mathbb{F}_2$ that consists of random rows $\hat{H}_i=(\hat{h}_{i1},\dots,\hat{h}_{in})$ for $1\le i \le m$, such that $\Pr(\hat{h}_{ij}=1)=\frac{1}{2}$ for  $1\le j\le n$.
Finally, let $\xi_k(n-s)=\Pr(\text{Rank}(\hat{H})=n-s)$ the probability that $\text{Rank}(\hat{H})=n-s$, where $k=m-n$ (or $k=\gamma n$).  

Let us  introduce Kovaleko's rank-distribution of  $\hat{H}$.  It is shown in \cite{kv1,cc,rg} by Kovalenko that, for any fixed integers $k$ and $s$ with $l=k+s\ge 0$,
\begin{equation}
 \xi_k(n-s)=\frac{S(n-s,l)}{2^{ls}} \prod_{i=s+1}^{n}\left(1-\frac{1}{2^i} \right), \label{eq:krt}
\end{equation} 
where
\begin{equation}
  S(n-s,l)=  \sum_{i_{1}=0}^{n-s}2^{-i_{1}}\sum_{ i_{2}=i_{1}}^{n-s}2^{-i_{2}}\cdots \sum_{i_l=i_{l-1}}^{n-s}2^{-i_l}.  \label{eq:mkvsumr}
\end{equation}
Since $\lim_{n\to\infty}S(n-s,l)=\prod_{i=1}^{k+s}(1-\frac{1}{2^i})^{-1}$, it holds that
\begin{equation}
\lim_{n\to\infty} \xi_k(n-s) = \frac{1}{2^{s(k+s)}}
 \frac{ \prod_{i=s+1}^{\infty} (1-\frac{1}{2^i})}
      {\prod_{i=1}^{k+s} (1-\frac{1}{2^i})}.       \label{eq:qrd}
\end{equation} 
In fact,  the limit distribution above still holds when entries of $\hat{H}$ meet the density constraint
\begin{equation}
 \frac{\ln(n)+x}{n}\le \Pr(\hat{h}_{ij}\neq 0) \le 1-\frac{\ln(n)+x}{n}, \label{eq:lnnx}
\end{equation}
where $x\to \infty$ arbitrarily slowly.
The limit distribution,  however,  is not directly applicable to $H$ in System~(\ref{sys:ini}), because entries of $H$  may not follow the  constraint (\ref{eq:lnnx}).

In the following, we define KFRL and verify it as a lower bound for $1-\xi_k(n)=\Pr(\text{Rank}(\hat{H})<n)$.  We then define  KFRO  from  KFRL and verify it as a lower bound for MSO. 
 Foremost, notice  that the sequence $\{ S(n-s,l) \}_{n=s}^{\infty}$ is in fact increasing, therefore,
\begin{equation}
 S(n-s,l)\; \le \; \lim_{n\to\infty}S(n-s,l)=\prod_{i=1}^{k+s}\left(1-\frac{1}{2^i}\right)^{-1}. \label{eq:S}
\end{equation}
By Plugging in $s=0$ into (\ref{eq:krt}) and (\ref{eq:S}),  we have 
\begin{equation}
 1-\prod_{i=k+1}^{n}\left(1-\frac{1}{2^i} \right)\, \le \, 1-\xi_k(n).\label{eq:ffrl}
\end{equation}
With the left-hand side  above, where $k=\gamma n$, define
\begin{equation}
 K(1+\gamma,n)=1-\underbrace{\prod_{i=k+1}^{n} \left( 1- \frac{1}{2^i}\right)}_{g(k,n)},  \label{eq:kfrl}
\end{equation}
 and refer to as KFRL.  
For a given error-bound $\delta$ now, define
\begin{equation}
\gamma_K(\delta,n)= \min_{\gamma \ge 0}\{\gamma \,|\, K(1+\gamma,n)\le \delta  \},\label{eq:kfro} 
\end{equation}
and refer to as the KFRO with $\delta$.  Notice that KFRL is decreasing with respect to $\gamma$, and thus,  $K(1+\gamma,n)\le \delta$ for any $\gamma \ge \gamma_K(\delta,n)$.
Observe from (\ref{eq:kfrl}) that $g(k+1,n)=\left(1-\frac{1}{2^{k+1}}\right)g(k,n)$.
Hence by $g(0,n):=0.288788095066$ for $n\ge 50$,  $K(1+\gamma,n)$ can be computed explicitly by (\ref{eq:kfro}), and consequently, $\gamma_K(\delta,n)$ is obtainable from the graph of $K(1+\gamma,n)$.

The following proposition shall be conveniently used for upper bounds for $K(1+\gamma,n)$ and $\gamma_K(\delta,n)$, and for the proof of \lmref{lm:kfrl}.
\begin{prop}\label{prop:1}
Let $V=(v_1,\dots,v_n) \in\mathbb{F}_2^n$ be given with $|V|=k >0$, and let $W=(w_1,\dots,w_n) \in\mathbb{F}_2^n$ be a random vector  such that  $\Pr(w_i=1)=\frac{d}{n}$ for $1\le i\le n$.  Then
\begin{equation}
 \Pr(W\cdot V^T=0)=\frac{1+ (1-\frac{2d}{n})^k}{2}, \label{eq:xy}
\end{equation}
where  $W\cdot V^T=\sum_{i=1}^n w_iv_i$ over $\mathbb{F}_2$.
\end{prop}

\begin{IEEEproof}
From binomial expansions, we have
\begin{equation}
 \sum_{s\, \text{even}} \binom{k}{s}a^sb^{k-s} = \frac{(a+b)^k+(-a+b)^k}{2}. \label{eq:ab}
\end{equation}
Let $p_i=\Pr(w_i=1)$ for $1\le i\le n$.
Since $|V|=k$, assume without loss of generality that $v_i=1$ for $1\le i\le k$ and $v_i=0$ for $k+1\le i\le n$, so that  
$\Pr(W\cdot V^T=0)=\Pr\left(\sum_{i=1}^k w_i=0\right)$. 
Then since $\sum_{i=1}^k w_i=0$ iff. $w_i=1$ for even number of $i$'s, 
\begin{equation}
  \Pr\left(\sum_{i=1}^k w_i=0\right)=\sum_{s\,\text{even}} \binom{k}{s} \prod_{i\in I_s}p_i \prod_{i\notin I_s}(1-p_i)
\end{equation}
where $I_s\subset\{1,2,\dots,k\}$ with $|I_s|=s$. 
Hence by $p_i=\frac{d}{n}$ for $1\le i\le n$, we have
\begin{equation}
 \Pr(W\cdot V^T=0)=\sum_{s\, \text{even}} \binom{k}{s} \left(\frac{d}{n}\right)^{s}\left(1-\frac{d}{n}\right)^{k-s}. 
\end{equation}
Taking  $a=\frac{d}{n}$ and $b=1-\frac{d}{n}$ into (\ref{eq:ab}) verifies  (\ref{eq:xy}). 
\end{IEEEproof}

%The theorem blow  shows  $\gamma_K(\delta,n)\le \frac{1+\log_2(1/\delta)}{n}$.

\begin{thm}[Upper-Bound for $\gamma_K(\delta,n)$] \label{thm:lbgk} For a given error-bound $\delta$, let $k_\delta >0$ be an integer such that
\begin{equation}
 \frac{\log_2(1/\delta)}{n} \le \left( \gamma_\delta=\frac{k_{\delta}}{n} \right) \le \frac{1+\log_2(1/\delta)}{n}, \label{eq:logdelta}
\end{equation} 
i.e., $k_\delta=\min\{k\in\mathbb{Z}\:|\:2^{-k}\le \delta\}$.
It then follows that
\begin{equation}
\gamma_K(\delta,n)\le \frac{1+\log_2(1/\delta)}{n}. \label{eq:kfrolb}
\end{equation}
\end{thm}

\begin{IEEEproof}
Let $\hat{H}$ be an $m\times n$ binary random matrix with $m=n+k_\delta$ such that, for each row $\hat{H}_i=(\hat{h}_{i1},\dots,\hat{h}_{in})$, $\Pr(\hat{h}_{ij}=1)=\frac{1}{2}$ for $1\le j\le n$. 
By \propref{prop:1},  $\Pr(\hat{H}_i\cdot V^T=0)=\frac{1}{2}$ for $1\le i\le n$ and  $V\neq 0$.  Then since each $\hat{H}_i$ is independent of all other rows, 
\begin{equation}
\Pr(V\in\text{Ker}(\hat{H}))=\prod_{i=1}^{m} \Pr(\hat{H}_i\cdot V^T=0) = \frac{1}{2^{m}}.
\end{equation}
Note that $\text{Rank}(\hat{H})<n$ iff. $\hat{H}\cdot V^T=0$ for some $V\neq 0$, and there are  of total $2^{n}-1$ nonzero vectors in $\mathbb{F}_2^{n}$.   Therefore,
\begin{equation}
1-\xi_{k_{\delta}}(n)  \le \sum_{V\neq 0}  \Pr\left(V\in\text{Ker}(\hat{H})\right) \le \frac{2^n-1}{2^m}  \frac{1}{2^{k_{\delta}}} < \delta. \label{eq:lbkfl}
\end{equation}
Hence by (\ref{eq:ffrl}), $K(1+\gamma_\delta,n) < \delta$, and by the definition of $\gamma_K(\delta,n)$, $\gamma_K(\delta,n)\le \gamma_\delta$. The inequality   (\ref{eq:kfrolb}) is then clear by (\ref{eq:logdelta}).
\end{IEEEproof}

Although the authors of the paper are not able to provide any mathematical proofs,  experiments exhibited that $K(1+\gamma,n)$ and $2^{-\gamma n}$ are almost identical as $\delta$ decreases. Hence $\gamma_K(\delta, n)$ is in fact very close to  $\gamma_\delta=\frac{k_\delta}{n}$.
Notice that, since $\lim_{n\to \infty}\frac{1+\log_2(1/\delta)}{n} =0$ as long as $\delta \ge 2^{-n^c}$ for  $c<1$,   $\lim_{n\to \infty}\gamma_K(\delta,n) =0$ for such $\delta$ by \thmref{thm:lbgk}.

In the following lemma, we show that $K(1+\gamma,n)\le  P_{ML}^{err} (1+\gamma,n)$. 
As a consequence of the lemma,   we show in ~\thmref{thm:kfro}  that   $\gamma_K(\delta,n)\le \gamma_*(\delta,n)$.

\begin{lem}[KFRL as a lower-bound for DEP]\label{lm:kfrl}
Let $H$ be an $m\times n$ matrix of System~(\ref{sys:ini}), where $m=(1+\gamma)n$ with $\gamma \ge 0$. Then
\begin{equation}
K(1+\gamma,n)\le  P_{ML}^{err}(1+\gamma,n).  \label{eq:ffrl1}
\end{equation}
\end{lem}

\begin{IEEEproof}
 Let $k=\gamma n$, $m=(1+\gamma)n$, and  $\hat{H}$ an $m\times n$ binary random matrix such that $\Pr(\hat{h}_{ij}=1)=\frac{1}{2}$. 
 We first show that 
\begin{equation}
\Pr(\text{Rank}(H)=n) \; \le \; \Pr(\text{Rank}(\hat{H})=n). \label{eq:ubfr2}
\end{equation}
In LT codes,  each row $H_i$ of $H$ in \sysref{sys:ini}  follows the uniform probability $\Pr(h_{ij}=1)=\frac{d}{n}$ with $d\le \frac{n}{2}$, where $d=|H_i|$ with probability $\mu_d$ of the RSD $\mu(x)=\sum \mu_d x^d$.  
In LDPC codes,  $H$ of \sysref{sys:ini} is formed by randomly chosen $n=pN$ columns of the check matrix $M$.  
In both LT and LDPC codes, thus,  $\Pr(h_{ij}=1)\le \frac{1}{2}$ for $1\le j\le n$.  
Then  by \propref{prop:1},     $\Pr(\hat{H}_i\cdot V^T=0) \le \Pr(H_i\cdot V^T=0)$ for  $V\in\mathbb{F}_2^n$, and  this is true for every $1\le i\le m$.  
Therefore,  $\Pr(\hat{H}\cdot V^T=0) \le \Pr(H\cdot V^T=0)$, and in expectation sense,  $|\text{Ker}(\hat{H})| \le |\text{Ker}(H)|$, and hence,  the inequality (\ref{eq:ubfr2}) is verified.  
The inequality (\ref{eq:ffrl1}) is then clear  by the lower bound in (\ref{eq:ffrl}). 
\end{IEEEproof}

\begin{thm}[KFRO as a lower-bound for MSO]\label{thm:kfro} To solve \sysref{sys:ini} uniquely with a destined bound $P_{ML}^{err}(1+\gamma,n)\le\delta$, it should hold that
\begin{equation}
 \gamma_*(\delta,n) \ge   \gamma_K(\delta,n). \label{eq:mfo}
\end{equation}
To achieve $P_{ML}^{err}(1+\gamma,n)\le \delta$, therefore, the numbers of symbols that receivers should acquire is at least  $(1+\gamma_K(\delta, n))n$ for LT codes, and $\frac{R+\gamma_K(\delta,n)}{1+\gamma_K(\delta,n)}N$ for LDPC codes.
\end{thm}
\begin{IEEEproof}
 The inequality (\ref{eq:mfo}) is clear by \lmref{lm:kfrl} and by the definitions of $\gamma_*(\delta,n)$  and $\gamma_K(\delta,n)$ in  (\ref{eq:mo}) and (\ref{eq:kfro}), respectively.  
To achieve $P_{ML}^{err}(1+\gamma,n)\le \delta$ with  LT codes, the inequality (\ref{eq:mfo}) implies that the number of symbols of $\beta$, equivalently, the  row-dimension $m$ of $H$ in System~(\ref{sys:ini}), should be at least $(1+\gamma_K(\delta,n))n$. 
In case of LDPC codes, recall that  $m=(1-R)N$ and $n=pN$. 
 To achieve $P_{ML}^{err}(1+\gamma,n) < \delta$ with LDPC codes, hence,  (\ref{eq:mfo}) implies that $m\ge (1+\gamma_K(\delta,n))pN$. 
 In other words, the number of lost symbols, equivalently the column-dimension of $H$ in System~(\ref{sys:ini}) that is $n=pN$, should  be at most  $\frac{(1-R)N}{1+\gamma_K(\delta,n)}$ where $(1-R)N=m$. Therefore,   the number of acquired  symbols by receivers, i.e., $(1-p)N$, should be at least $\frac{\gamma_K(\delta,n)+R}{1+\gamma_K(\delta,n)}N$. 
\end{IEEEproof}

\begin{exmp}
Red curves in Fig.~\ref{fig:Perf} represent the KFRL $K(1+\gamma,n)$, where  $n=100$ for LT codes (top) and $n=p200$ for LDPC codes (bottom) with $0\le p\le \frac{1}{2}$.
When $\delta=10^{-4}$, for an example, $1+\gamma_K(10^{-4},n)\approx 1.14$ in both LT and LDPC codes. To verify $1.14$ with LDPC codes, use the conversions in (\ref{eq:gammap}) with $p_K\approx 0.43$ in the bottom figure. This implies that by \lmref{lm:kfrl}, since $K(1+\gamma, n)\ge 10^{-4}$ for $1+\gamma \le 1.14$, the DEP of both LT and LDPC codes  can not  be better than $10^{-4}$,  i.e., $P_{ML}^{err}(1+\gamma,n)\ge 10^{-4}$ for $\gamma \le 0.14$.   
Again by \thmref{thm:kfro}, to achieve  $P_{ML}^{err}(1+\gamma,100) \le 10^{-4}$ with LT codes, the minimum overhead  $\gamma_*(10^{-4},100)$ should be larger than  $0.14$, i.e., $\gamma_*(10^{-4},n) \ge 0.14$.  Analogously, to achieve $P_{ML}^{err}(1+\gamma,n) \le 10^{-4}$ with the LDPC codes, where $n=p200$, the maximum tolerable loss rate $p^*= \frac{0.5}{1+\gamma_*(10^{-4},n)}$ (use the conversion in (\ref{eq:gammap})),  should be less than $p_K=\frac{0.5}{1+\gamma_K(10^{-4},n)}\approx 0.43$, i.e.,  $p^*\le 0.43$.  

Another thing should be noticed is that,  as mentioned earlier,  the two curves $K(1+\gamma,n)$ and $2^{-\gamma n}$ in the top figure are almost identical as $\delta$ decreases.  In this respect, $\gamma_K(10^{-4},100)\approx \frac{k_\delta}{100}$, where $k_\delta$ is the smallest integer $k$ such that $2^{-k}\le 10^{-4}$. 
It is not hard to see by direct computation that $k_\delta=14$ for $\delta=10^{-4}$ and  $\gamma_\delta \approx \frac{14}{100} =0.14$, that is precisely the $\gamma_K(10^{-4},100)$.   \hfill 
\end{exmp}

\section{Experimental Results with LT and LDPC Codes  \label{simulations}}

\begin{table}[ht]
\begin{center}
\begin{tabular}{|c|l|}
 \hline%\hline
                 & $(\mu_d)_{d=1}^{5}=(0.012,0.482,0.153,0.082,0.047)$ \\
                 & $(\mu_d)_{d=6}^{10}=(0.035,0.024,0.023,0.012,0.012)$ \\   
        $\mu(x)$ & $\mu_{25}=0.059$, $\mu_{35}=0.059$ \\ \hline
    $\bar\mu(x)$ & Normalization of $\mu(x) +(0.17)x^{50}$\\
    \hline
 \hline%\hline
                 & $(\rho_d)_{d=2}^8=(0.46, 0.32, 0.021, 0.06, 0.04,0.025)$\\   
       $\rho(x)$ & $\rho_{9}=0.01,\rho_{19}=0.02,\rho_{20}=0.05$\\
                 \hline
                % \\
   %$\bar\rho(x)$ & Normalization of $\rho(x)+(0.17)x^{50}$\\
   %\hline 
\end{tabular}
\end{center}
\caption{The row-degree distributions $\mu(x)$ and $\bar{\mu}(x)$ for LT codes (top), and the column-degree distributions $\rho(x)$  for LDPC codes (bottom).}\label{tab:rho}
\end{table}

\begin{figure}[ht]
 \centerline{ \fbox{\epsfig{file=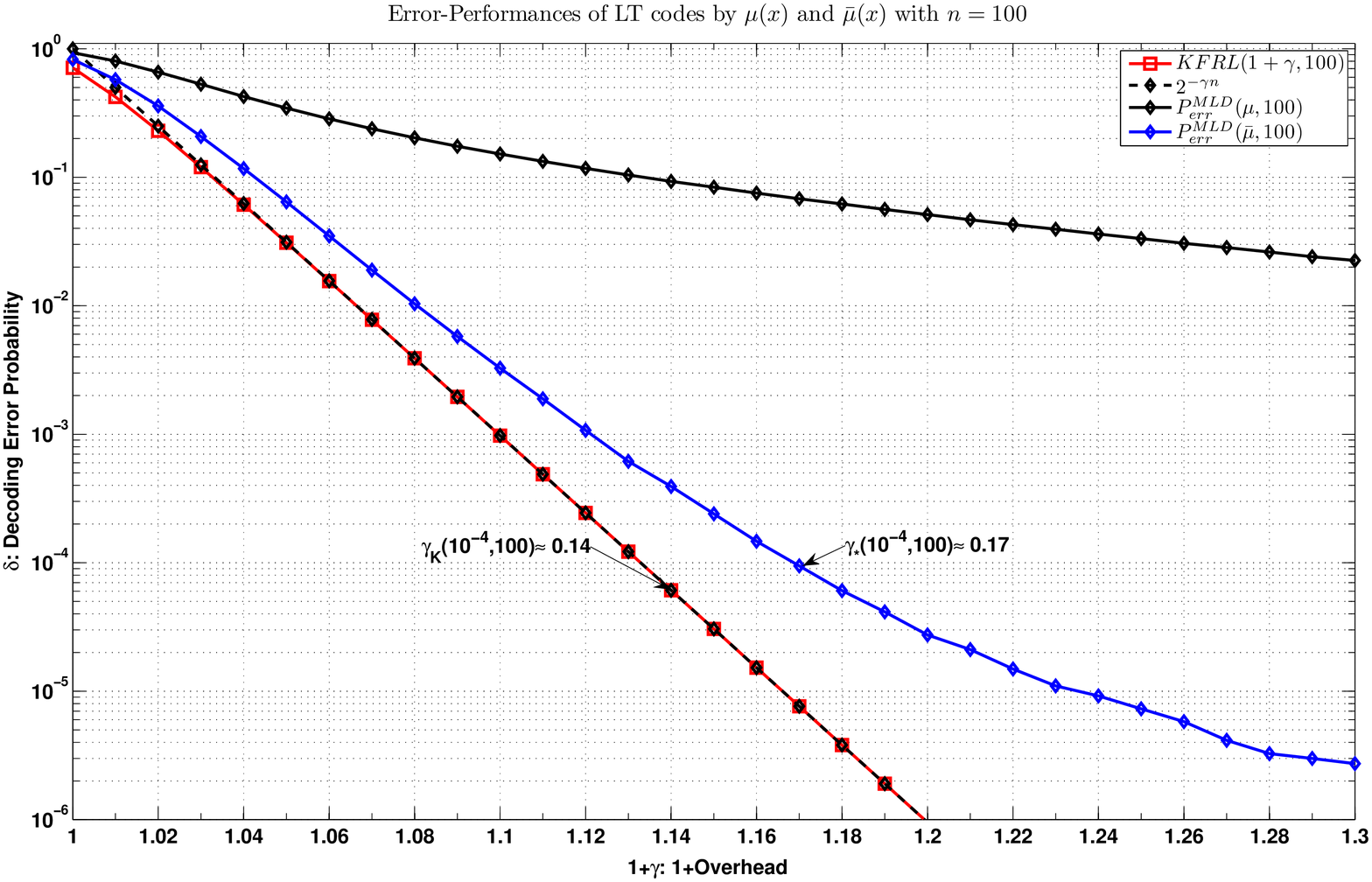, width=16cm,height=10cm}}}
 \vskip 10pt
 \centerline{ \fbox{\epsfig{file=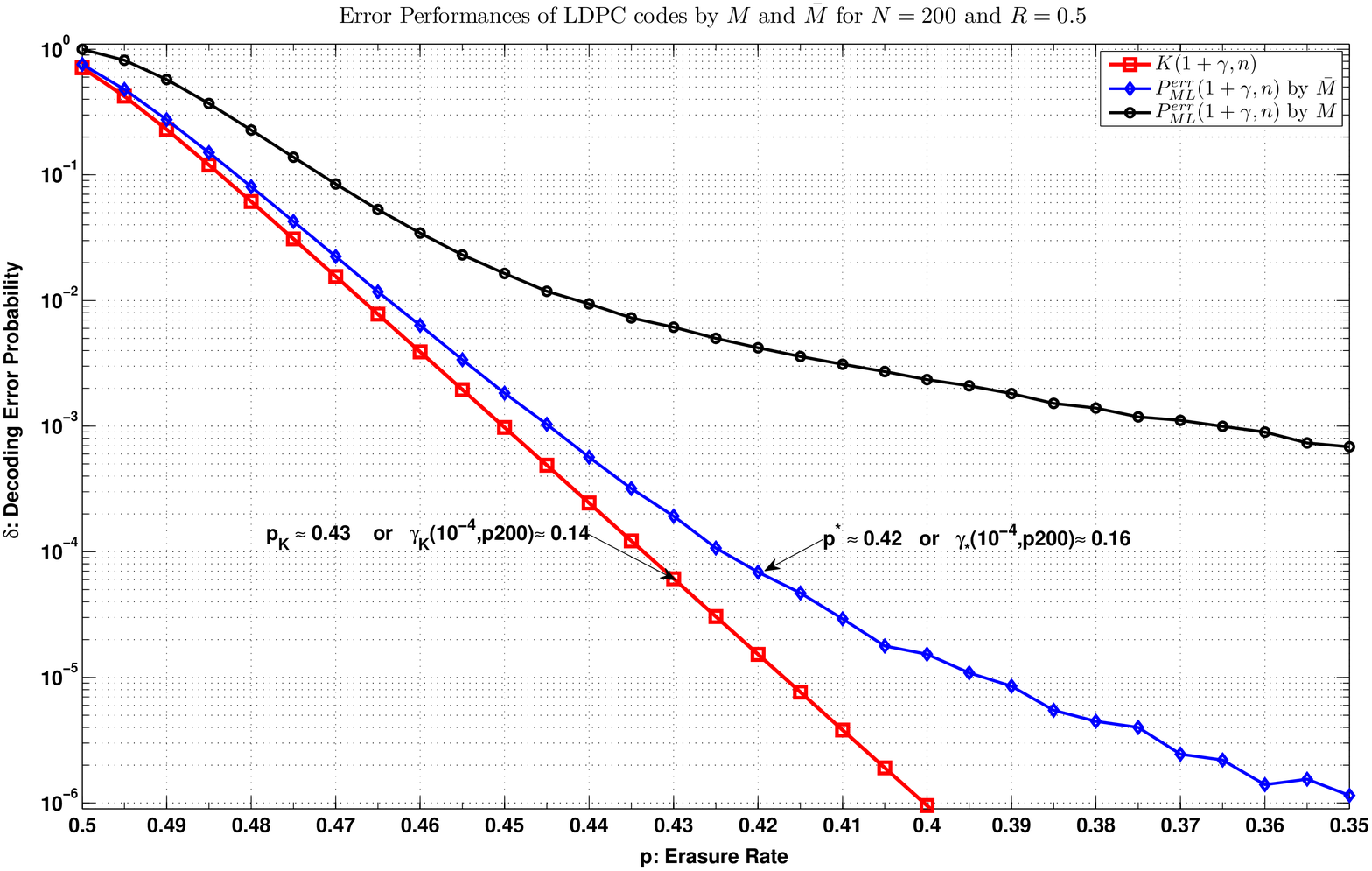, width=16cm,height=10cm}}}
 \centering 
\caption{ Top figure shows the error-performance   of  LT codes by $\mu(x)$ (black) and $\bar{\mu}(x)$ (blue) in DEP vs. overhead.  Bottom figure shows  the error-performance of LDPC codes by $M$ (black) and $\bar{M}$ (blue) in DEP vs. erasure rate, where $p=\frac{1-R}{1+\gamma}$. 
\label{fig:Perf}
} 
\end{figure}

In this section, we provide experimental results  which  show the viability that  both LT and LDPC codes may achieve the error-performances in $P_{ML}^{err}(1+\gamma,n)$ and $\gamma_*(\delta,n)$ that are close to  $K(1+\gamma,n)$ and  $\gamma_K(\delta,n)$, respectively, when enough number of dense rows or columns are supplemented to $H$ in System~(\ref{sys:ini}).  
Codes for experiments are arranged as following.
 For LDPC codes, two check matrices of block dimension $100\times 200$ (thus $R=\frac{1}{2})$, say $M$ and $\bar{M}$, were arranged by using PEG algorithm in \cite{peg}:  $M$ was generated with the column-degree distribution $\rho(x)$ in \tabref{tab:rho} and $\bar{M}$ was generated by supplementing $15$ random rows of degree $\frac{N}{2}=100$ to a check matrix of dimension $85\times 200$ arranged with $\rho(x)$.
 For LT codes, two row-degree distributions $\mu(x)$ and $\bar\mu(x)$  in \tabref{tab:rho} were used for constructing codes of block length $n=100$. 
 
In \figref{fig:Perf}, curves represent $K(1+\gamma,n)$'s (red ones) and $P_{ML}^{err}(1+\gamma,n)$'s of LT and LDPC codes (blue and black ones), where $n=100$ for LT and $n=p200$ for LDPC codes with $0\le p\le 0.5$.  At each point of the DEP curves, the value of $P_{ML}^{err}(1+\gamma,n)$ is estimated by the fraction of the number of rank-deficient cases of  $m\times n$ matrices $H$ with $m=(1+\gamma)n$ (or the fraction of decoding failure cases of \sysref{sys:ini}) based on  more than $10^6$ random constructions of $(H,\beta)$ of \sysref{sys:ini}.  
The the Separated MLDA in \cite{ourlt,shortlt} was used to check the rank-deficiency.

It can be clearly seen  from the figure that, when check matrices of codes are constructed with  $\mu(x)$ and $\rho(x)$ that have no dense fractions (i.e. $\mu_{50}=\rho_{100}=0$), their DEP (black ones) never drop to the error-bounds, $\delta=10^{-2}$ with LT codes and $\delta=10^{-3}$ with LDPC codes.  These error-flooring phenomena are obviously due to  the deficient cases of $H$, i.e., $\eta=\dim \text{Ker}(H) >0$ that occur sporadically for large $\gamma$. Most of the deficient cases, however, $\eta$ is merely one or two for large $\gamma$ (small $p$ for LDPC codes). This small deficiency can be readily removed by supplementing a fraction of dense rows.   To improve their DEP, we altered $\mu(x)$ of the LT code into $\bar\mu(x)$ by supplementing the dense fraction  $\mu_{50}=0.17$ (thus $\bar\mu_{50}\approx 0.15$), and the check matrix $M$ was redesigned to $\bar{M}$ by supplementing $15$ random rows of degree $100$ as stated before. Thus, $H$ in \sysref{sys:ini} by  $\bar\mu(x)$ and $\bar{M}$ can have enough number of dense rows.  By doing so, the altered codes achieved their DEP curves (blue ones) and MSO $\gamma_*(\delta,n)$  that are close to the lower bounds KFRL and KFRO for  $\delta \le 10^{-4}$, respectively.   

It is interesting to note that $K(1+\gamma,n)$ is very close to $2^{-\gamma n}$ for small $\delta$.   In this case, $\gamma_K(\delta,n)$ can be understood as  the integer $k_\delta$ such that $\log_2(1/\delta)\le k_\delta \le 1+\log_2(1/\delta)$, i.e., $\gamma_K(\delta,n):=\frac{k_\delta}{n}$.

Although we do not present experimental evidences, supplementing  about $15$ percent of dense rows to $H$ of \sysref{sys:ini} does not degrade the computational complexity of solving \sysref{sys:ini} seriously.  For an example, with the LT codes generated by the $\bar\mu(x)$,  the number of symbol additions on $\beta$ of \sysref{sys:ini} to compute the  solution of the system under the Separated MLDA is within $1,100$ (that is $11n$). Similarly with the LDPC codes by $\bar{M}$, the number of symbol addition on $\beta$ is within $1,600$ (that is $8N$).

\section{Summary \label{summary}}
We presented that Kolvalenko's full-rank limit and its overhead are tight lower bounds for decoding error probability and minimum stable overheads, respectively,  of LT and LDPC codes.  We also provided experimental evidences which show the viability that, when enough number of dense rows are supplemented to check matrices, both LT and LDPC codes may achieve the code performances in decoding error probability  and minimum stable overheads that are close to Kovalenko's full-rank limit and its overhead, respectively.

%As further research projects, the authors of the paper are investigating for optimal  dense fractions for the supplementation (see \cite{dltd} for LT codes and \cite{dldpcd} for LDPC codes). 
%The authors conjecture that KFRL is a very tight lower-bound for the error-performance of LT and LDPC codes over BEC in the following sense: For any $n$, they may be designed to achieve their $\text{DFR}(1+\gamma,n)$s that are very close to $K(1+\gamma,n)$ by supplementing a certain number of dense rows to check matrices, and the number is relatively small compared to $n$.

% \fi
\IEEEpeerreviewmaketitle

\end{document}